\documentclass[twocolumn]{article}

\usepackage{amsmath, amssymb}
\usepackage{graphicx}
\usepackage{physics}
\usepackage{bm}
\usepackage{authblk}
\usepackage{float}
\usepackage{siunitx}
\usepackage{tabularx}
\usepackage[dvipsnames]{xcolor}
\usepackage[switch]{lineno}

\title{Methods for traceable scanning magnetometry using single nitrogen vacancy centers in diamond: determining orientation, distance and localization }

\author[1]{Nikhita Khera}
\author[1]{Ephraim Spindler}
\author[1]{Yanis Abdedou}
\author[3]{Marcel Gasser}
\author[2]{Sandra Wolff}
\author[2]{Bert Lägel}
\author[3]{Robert Frömter}
\author[1]{Mathias Weiler}
\author[3]{Mathias Kläui}
\author[1]{Elke Neu}

\affil[1]{Department of Physics and Research Center OPTIMAS, RPTU University of Kaiserslautern-Landau, 67663 Kaiserslautern, Germany}
\affil[2]{Nano Structuring Center (NSC), RPTU University of Kaiserslautern-Landau, 67663 Kaiserslautern, Germany}
\affil[3]{Institut für Physik, Johannes Gutenberg-Universität Mainz, 55099, Mainz, Germany}

\date{\today}

\begin{document}
\maketitle

\begin{abstract}

Individual, scannable nitrogen vacancy (NV) centers in single crystal diamond nanostructures enable nanoscale, quantitative imaging of magnetic stray fields. Nevertheless, important parameters like distance between the NV center and the sample and the orientation of the NV high symmetry axis are often not known precisely and enter data evaluation as free fitting parameters. We here use scanning NV imaging on micro-patterned,  perpendicularly magnetized stripes and discs. From these measurements, we directly infer NV–sample distance $d_{\mathrm{NV}}$ and the NV's azimuthal orientation without the need for an external vector magnet control. We determine $d_{\mathrm{NV}} =$\SI{31.5}{\nano\meter}, while we infer the azimuthal orientation with a precision of \ang{3}.  We additionally employ commercially available silicon needles to image the apex topography of our diamond nanostructures to detect surface contamination. Simultaneously, monitoring NV fluorescence as a function of the needle's position allows us to estimate the  lateral placement of the NV inside the diamond nanostructure.  
\end{abstract}

\section{Introduction}
\label{sec:intro}

\begin{figure*}[!t]
    \centering
    \includegraphics[width=\textwidth]{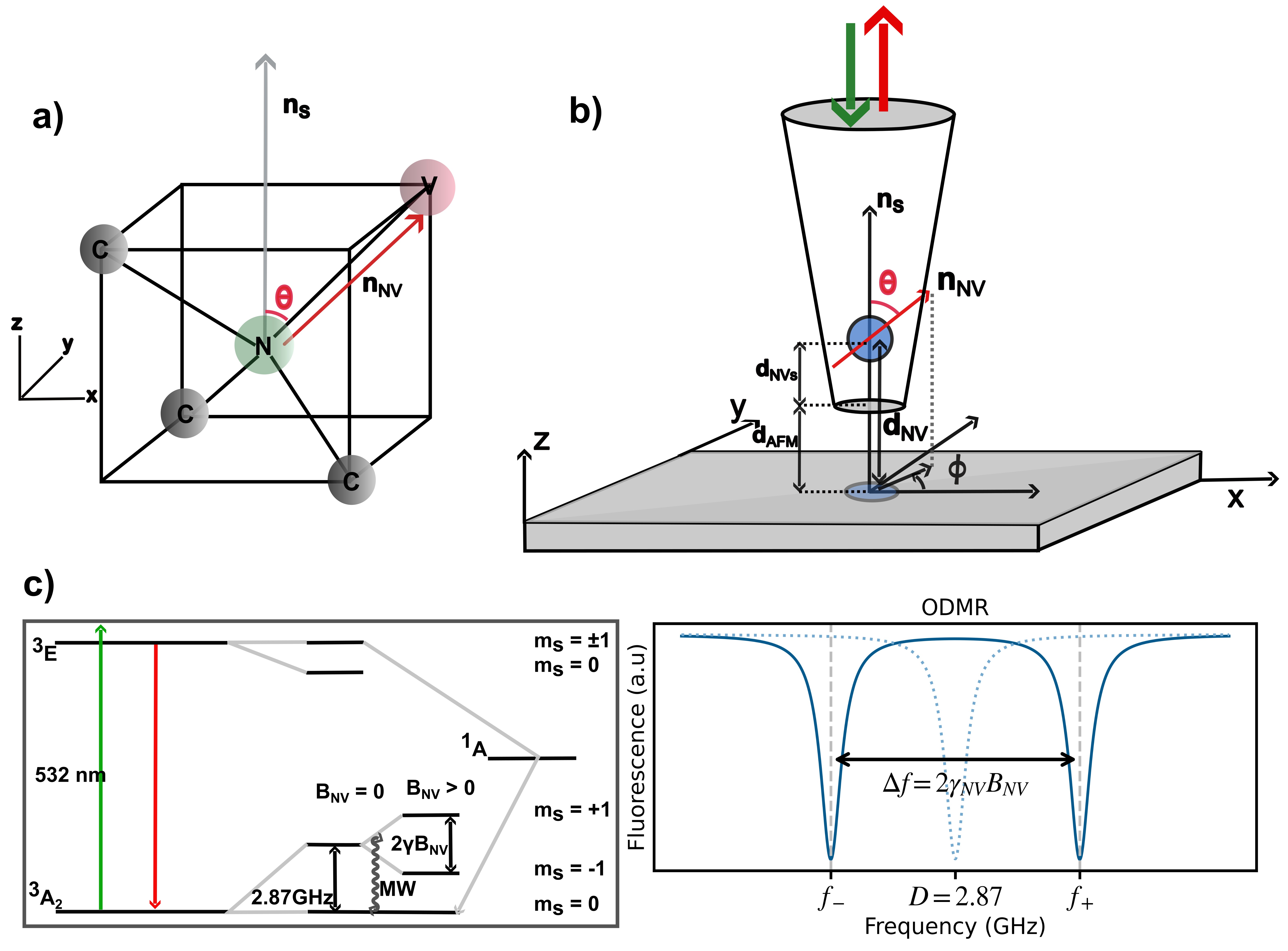}
    \caption{\textbf{Nitrogen-Vacancy (NV) center geometry and sensing principles.} 
    (a) Unit cell of the diamond lattice highlighting the NV center orientation. The Nitrogen (N) and Vacancy (V) define the high symmetry axis $\mathbf{n}_{\text{NV}}$, forming a polar angle $\theta$ with respect to the surface normal, $n_s$ along the $z$-axis. 
    (b) Schematic of the scanning NV microscopy setup. The NV-containing diamond nanopillar probe is scanned close to the sample surface. The NV has a stand-off distance $d_{\text{NV}}$ above the sample surface. The magnetic field projection depends on both the polar angle $\theta$ and the azimuthal angle $\phi$. 
    (c) The energy level diagram of the NV (left) and the Optically Detected Magnetic Resonance (ODMR) spectrum (right). The $m_s$ levels in the ground state triplet $^{3}A_2$ undergoe Zeeman splitting proportional to the magnetic field projection $B_\mathrm{NV}$, where $\Delta f = 2\gamma_{\mathrm{NV}} B_\mathrm{NV}$.}
    \label{fig:NVSchematic}
\end{figure*}

Individual, scannable nitrogen vacancy (NV) centers in single crystal diamond nanostructures enable nanoscale, quantitative, non-invasive imaging of magnetic stray fields. Typically, scanning NV microscopy (SNVM) systems, home built or commercial, host an NV center in  a conical diamond nanopillar probe that simultaneously serves as a tip for AFM feedback in scanning and as nano-waveguide for the NV's fluorescence. At the heart of all NV magnetometry techniques is the NV's optically readable electronic spin ~\cite{Rondin2014}. A local magnetic field shifts the spin's resonance frequencies via the Zeeman effect, and this shift can be read out optically. The measured quantity is $B_{\mathrm{NV}}$, the projection of the local magnetic field onto the NV's high symmetry axis $\mathbf{n}_{\mathrm{NV}}$ [Figure~\ref{fig:NVSchematic}(a)].
Over the past decade, SNVM has been  applied to investigate a broad range of systems, e.g.\ spin textures in antiferromagnets~\cite{Gross2017}, magnetic textures in thin films~\cite{Tetienne2015},  mapping current distributions in electronic devices~\cite{Chang2017}, and novel magnetic phenomena like frustrated magnets (spin-ice)~\cite{Spindler2025}. Typical sensitivities for magnetometry with single NV centers are in the order of $\frac{\mu T}{\sqrt{Hz}}$ (the standard figure of merit for magnetometry, independent of integration time and limited by photon shot noise) with spatial resolution on the order of \SI{50}{\nano\meter}. More recently, techniques beyond static observation of Zeeman shifts have been developed: NV spin relaxometry has imaged antiferromagnetic textures~\cite{Finco2021} and external paramagnetic spins~\cite{Pelliccione2014}, while coherent spin-wave transport has also been imaged~\cite{Simon2022} using scanning NV techniques. 

Despite this high potential of SNVM, a common drawback in quantitative analysis is that the distance between the NV center and the sample, $d_{\mathrm{NV}}$, and the orientation of the NV axis with respect to the laboratory frame (angles $\theta$ and $\phi$) is not known precisely. $d_{\mathrm{NV}}$, $\theta$ and $\phi$, however, directly enter the reconstruction of magnetic fields in the sample plane as well as the sample's magnetization from  $B_{\mathrm{NV}}$ measured at the NV's location. 

Typically, NV centers are created via nitrogen ion-implantation allowing to place them in a depth $d_{\mathrm{NV_S}}$ below the diamond surface. However,  straggle of the implanted ions leads to a significant uncertainty of $d_{\mathrm{NV_S}}$: for an implantation energy of \SI{6}{\kilo\electronvolt} (the energy used to fabricate the probes in this work), $d_{\mathrm{NV_S}}$ is $\approx$ \SI{10}{\nano\meter},  but straggle as large as \SI{10}{\nano\meter} - \SI{20}{\nano\meter}~\cite{Maletinsky2012, Gross2017, Toyli2010} has been reported. Thus $d_{\mathrm{NV_S}}$ varies for individual NVs. Additionally, technical constraints (non-optimal AFM feedback, probe contamination, tilt of diamond nanostructure, sample roughness) induce an additional stand-off distance $d_{\mathrm{AFM}}$, where $d_{\mathrm{NV}}=d_{\mathrm{NV_S}}+d_{\mathrm{AFM}}$ as depicted in Figure~\ref{fig:NVSchematic}(b). As a trend, optimized AFM operation during scanning reduces $d_{\mathrm{AFM}}$~\cite{Xu2025} and a smaller $d_{\mathrm{AFM}}$ is reached in cryogenic SNVM under vacuum as compared to ambient condition SNVM~\cite{Thiel2016}. Even for a given probe, $d_{\mathrm{AFM}}$ and consequently $d_{\mathrm{NV}}$ can change over time due to probe contamination. Consequently, calibrating $d_{\mathrm{NV}}$ is required not only for each new probe, but preferably before each new set of measurements. 

Calibrating the orientation of the NV high symmetry axis has different requirements: typically, the single crystal diamond's orientation from which the nanostructures are manufactured sets $\theta$. In our case, the nanopillar apex is a (100) plane of diamond fixing $\theta=$\ang{54.7}. However, the in-plane azimuthal angle, $\phi$, is unknown as individual NV centers statistically align along any of the four crystallographically equivalent $\langle 111 \rangle$ directions in diamond. In many setups, $\phi$ is determined using defined externally applied magnetic fields (vector magnets) which are however not available in all setups. 

While $d_{\mathrm{NV}}$, $\theta$ and $\phi$ fully characterize the NV's spin properties, the lateral placement (x-y-plane) in the nanopillar will influence how efficiently the NV's fluorescence couples into the photonic mode of the nanopillar~\cite{Fuchs2018}. Consequently, this influences the detected fluorescence rate from a single NV center, the optical spin-readout and consequently the sensitivity. The lateral placement moreover becomes highly relevant when the magnetic signal is to be correlated to topographic data of the sample under investigation. We here present a method, in which the diamond nanopillar is scanned over an ultra-sharp (radius of curvature $\leq$~\SI{10}{\nano\meter}) commercially available silicon needle (``inverse AFM''). We monitor the NV's fluorescence and infer the lateral localization of the NV in the nanopillar from that. Moreover, topography imaging using the ultra-sharp needles efficiently reveals probe contamination.

In this work, we detail a calibration procedure performed directly in our SNVM setup, using simple and reproducible sample geometries (patterned discs and stripes) without relying on additional hardware. We demonstrate a robust, three-step protocol using  Perpendicular Magnetic Anisotropy (PMA) microstructures and commercially available silicon needles. We combine line-cuts across PMA stripe edges with field maps of micron-sized discs of the same PMA material to deduce $d_{\mathrm{NV}}$ and $\phi$. Calibrating $d_{\mathrm{NV}}$ can be readily implemented as a standard step prior to any measurement. Inverse AFM using silicon needles reveals the contamination state of the probe and indicates the lateral localization of the NV.

\section{Methods and Materials}

\subsection{NV Basics}
\label{sec:model}

Evaluating optically detected magnetic resonance spectra of a single NV gives $B_{\mathrm{NV}}$, the projection of the local magnetic field onto the NV's high symmetry axis $\mathbf{n}_{\mathrm{NV}}$ [see also \ref{fig:NVSchematic} a)].  Thus, for each position of the probe

\begin{align}
B_{\mathrm{NV}} = \mathbf{n}_{\mathrm{NV}} \cdot \mathbf{B}
\label{eq:Bnv}
\end{align}
holds, where $\mathbf{B}$ is the magnetic field generated by the sample at the location of the NV center.

We define the sample coordinate system such that the $z$-axis corresponds to the direction of the surface normal, $n_{s}$ (pointing outward from the sample towards the nanopillar probe), while $x$ and $y$ lie in the sample plane. We fix the $x$-direction along the scan direction of our SNVM.

The orientation of $\mathbf{n}_{\mathrm{NV}}$ can then be parametrized as

\begin{align}
\mathbf{n}_{\mathrm{NV}} = 
(\sin\theta \cos\phi,\; \sin\theta \sin\phi,\; \cos\theta)
\label{eq:n_nv}
\end{align}

where $\theta$ is the polar angle with respect to the surface normal, $n_{s}$, and $\phi$ is the in-plane azimuthal angle. For our (100)-oriented diamond nanopillar probes, $\theta$ is \ang{54.7}, while we will determine $\phi$ (Section~\ref{sec:results} and Figure~\ref{fig:QNAMIphi} give more details on the probe geometry).

Experimentally, we obtain $|B_{\mathrm{NV}}|$  from the Zeeman splitting, given by the frequency of the $m_\mathrm{s} = 0 \to \pm 1$ spin transitions,
\begin{align}
\Delta f = 2\gamma_{\mathrm{NV}} |B_{\mathrm{NV}}|
\label{eq:deltaf}
\end{align}

where $\gamma_{\mathrm{NV}} \approx$ \SI{28}{\giga\hertz\per\tesla}  is the gyromagnetic ratio of the $NV$ electron spin. While the splitting gives the magnitude of the field projection along the NV axis, $|B_{\mathrm{NV}}|$, we note that when continuously tracking e.g.\ the  $m_\mathrm{s} = 0 \to 1$ transition, $B_{\mathrm{NV}}$ can be observed.  In this work, however, we only work with $|B_{\mathrm{NV}}|$, calculated from the ODMR spectra measured at different points. More generally, the transverse component can also be estimated from the contrast of the ODMR resonances~\cite{Spindler2025}. 

This relation forms the basis of the calibration procedure described below: given a model for the stray magnetic field $\mathbf{B}(x,y,z)$ generated by a reference structure, we use the measured $B_{\mathrm{NV}}$ to extract $d_{\mathrm{NV}}$ and $\phi$.

\subsection{Analytic Expressions for Stray Fields from PMA Microstructures}
\label{sec:pma}

\begin{figure*}[t]
    \centering
    \includegraphics[width=\textwidth]{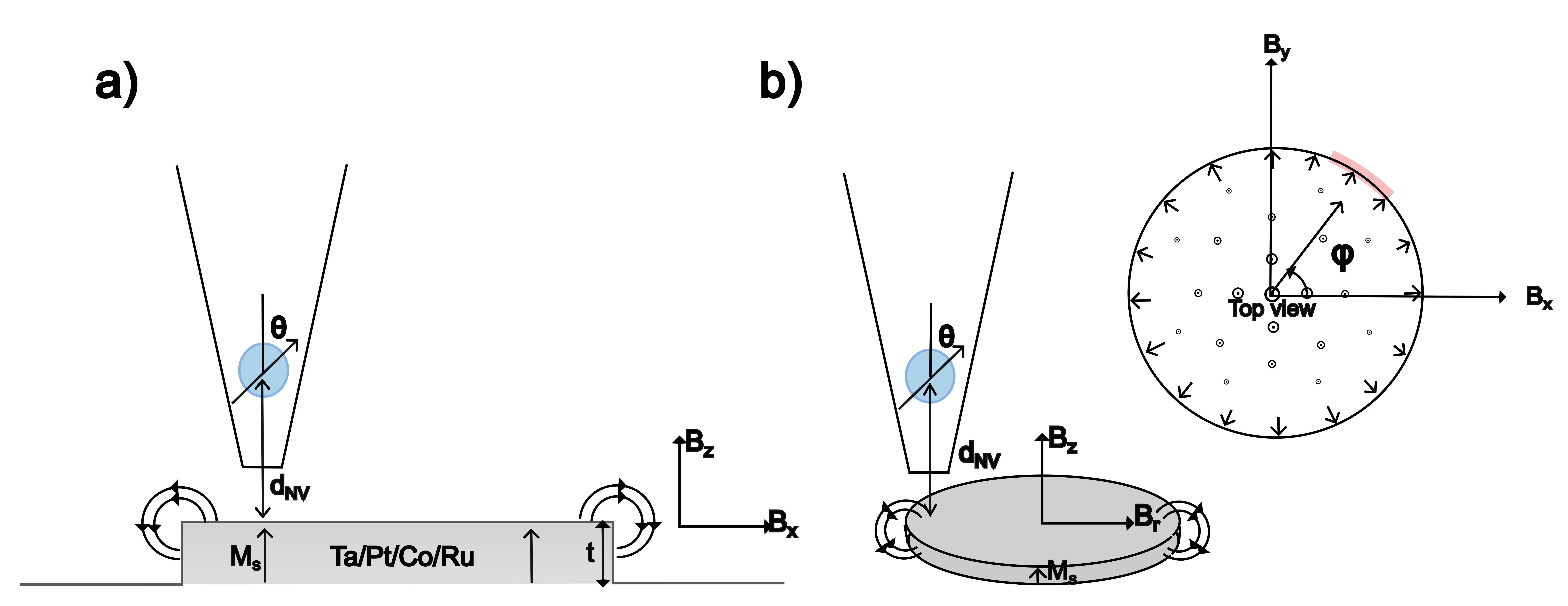}
    \caption{\textbf{(a)} Schematic illustrating the NV center in the probe, at a stand-off distance $d_{\mathrm{NV}}$ above the stripe edge. The out of plane magnetization $\mathbf{M_s}$, produces a stray field along the edge.
    \textbf{(b)} Schematic of the scanning NV nanopillar probe above a PMA disc, shown in side view (left) and top view (right). The out-of-plane magnetization $\mathbf{M}$ produces a radial stray field at the disc edge (arrows). The angular position of the field extremum along the edge (highlighted arc) encodes the azimuth $\phi$ of the NV axis, measured from the $+x$ axis.}
    \label{fig:models}
\end{figure*}

We model the stray magnetic fields generated by two distinct geometries of Perpendicular Magnetic Anisotropy (PMA) thin-film structures (for details on the PMA material see Section \ref{sec:methods}). In the thin-film limit, the magnetic thickness $t$ is much smaller than the sensing height $d_{\mathrm{NV}}$. In this limit, we treat the magnetic stripe as a surface distribution of magnetic dipoles with an effective area magnetization of $M_\mathrm{s} t$ [Figure~\ref{fig:models}].

\paragraph{PMA Stripe} 
For a semi-infinite magnetic stripe extending along the $y$-axis with an edge at $x=x_0$, the stray field is uniform along the stripe length, effectively reducing the problem to two dimensions ($B_y \approx 0$). At a vertical distance $d_{\mathrm{NV}}$ from the surface of the PMA film, the field components are given by~\cite{Hogen2023, Hingant2015}:

\begin{align}
B_x(x) &= \frac{A}{2\pi}\frac{d_\text{NV}}{(x-x_0)^2 + d_\text{NV}^2}, 
\label{eq:bx_stripe}
\\
B_z(x) &= -\frac{A}{2\pi}\frac{x-x_0}{(x-x_0)^2 + d_\text{NV}^2}, \label{eq:bz_stripe}
\end{align}
where $A = \mu_0 M_s t$. Our NV sensor therefore measures $B_{\mathrm{NV}}(x) = n_x B_x(x) + n_z B_z(x) + B_0$.  where $B_0$ is a small uniform bias field present throughout all measurements presented here. $B_0$ is measured independently via ODMR with the probe positioned away from any sample, prior to the measurements. As with all ODMR measurements, this yields the magnitude of the projection of $B_0$ on the NV-axis; we accordingly adopt the convention $B_0 \geq 0$, and constrain it as such in the fits described below. This field lifts the zero-field degeneracy of the  $m_s \pm 1$ spin sublevels, allowing a resolvable ODMR splitting everywhere on the scan -- including away from the stripe edge or disc. Also, since any change in field lies entirely in the $xz$-plane ($B_y = 0$), the stripe sample geometry is an ideal tool to infer $d_{\mathrm{NV}}$, which remains``blind'' to the full three-dimensional orientation ($\phi$)


\paragraph{PMA disc}
To deduce  $\phi$, we use a disc of radius $R$ patterned from the same PMA film. Unlike the stripe, the disc generates a two-dimensional stray field with both in-plane and out-of-plane components. Since its thickness is negligible ($t \ll d_{\mathrm{NV}}$,  details see Section~\ref{sec:methods})), we approximate the stray field by the field of a current carrying loop of radius $R$ carrying current $I = \sigma M_s t$, where $\sigma = \pm 1$ is the magnetization direction. The loop's magnetic field components can be expressed as :

\begin{align}
B_z(r,z) &= P\left[K(k^2) + \frac{R^2-r^2-z^2}{q}\,E(k^2)\right], 
\label{eq:bz_loop}
\\
B_r(r,z) &=   P\,\frac{z}{r}\left[-K(k^2)
                  + \frac{R^2+r^2+z^2}{q}\,E(k^2)\right]
\label{eq:br_loop}   
\end{align}

where $K$ and $E$ are the complete elliptical integral of the first and second kind, $P = \mu_0 I / \bigl(2\pi\sqrt{(R+r)^2+z^2}\,\bigr)$,  $q = (R-r)^2+z^2$ and $k^2 = 4Rr/\left[(R+r)^2+z^2\right]$~\cite{Jackson1998, Simpson2001}.

In polar coordinates $(r, \alpha)$ centered on the disc, the  measured field $B_{\mathrm{NV, \phi}}(r, \alpha)$ reads 

\begin{align}
    B_{\mathrm{NV} \phi}(r,\alpha) =     B_0 + \cos\theta \, B_z(r, d_{\mathrm{NV}}) \notag\\
 + \sin\theta \, B_r(r, d_{\mathrm{NV}})\cos(\alpha-\phi),
\label{eq:projection}
\end{align}

with $B_0$ being the small bias field introduced in the previous section.

\begin{table*}[!t]
\centering
\caption{Fabrication parameters for the calibration structures.}
\begin{tabularx}{\textwidth}{lXX}
\hline
 & \textbf{Stripe} & \textbf{Disc} \\
\hline
Resist & AZ~1512\,HS (positive); Spin coating: 4000\,rpm for \SI{30}{\second}; Soft bake: \SI{60}{\second} [\SI{100}{\degreeCelsius}]; Thickness: $\sim$ \SI{1.2}{\micro\meter} & ma-N~2403 (negative) ; Spin coating: 4000\,rpm for \SI{30}{\second}; Soft bake: \SI{60}{\second} [\SI{100}{\degreeCelsius}]; Thickness: $\sim$\SI{200}{\nano\meter}\\
Lithography & Direct laser writing (Microtech LW405D); Dose: \SI{156}{\milli\joule} & Electron-beam (Raith VOYAGER); Voltage: \SI{50}{\kilo\volt} \\
Developer & AZ~726 MIF (2.38\% TMAH), \SI{45}{\second} & AZ~726 MIF (2.38\% TMAH), \SI{45}{\second} \\
Etching & Ar ion-beam etching,  \SI{60}{\second} (beam voltage:  \SI{500}{\volt}) & Ar ion-beam etching,  \SI{60}{\second} (beam voltage:  \SI{500}{\volt})  \\
Resist removal & DMSO, Ultrasonication  \SI{10}{\minute}, Rinse with Distilled Water & DMSO, Ultrasonication  \SI{10}{\minute} at  \SI{70}{\degreeCelsius} , Rinse Distilled Water \\
\hline
\end{tabularx}
\label{tab:Fabrication}
\end{table*}

We note that the sign of $B_0$ and $\sigma$ cannot both be determined in absolute terms from the magnitude map alone; resolving this requires external knowledge of $B_0$'s sign. As described in Sec.~\ref{sec:pma}, $B_0$ is independently measured and constrained to be positive in the fit, which is what allows $\sigma$ to be determined relative to this fixed reference.

In contrast to the stripe geometry, the radial stray field of the disc leads to an azimuth dependent $B_{\mathrm{NV}}$. Consequently, the measured $B_{\mathrm{NV}}$ varies along the disc edge. Since $\Delta f$ (Eq.~\eqref{eq:deltaf}) depends only on the magnitude of the field projection, the experimentally accessible quantity is $|B_{\mathrm{NV}}|$. The measured map therefore shows two maxima on the disc edge, \ang{180} apart: the true one at $\alpha=\phi$, and a dimmer one arising from the field minimum being projected to positive values as we measure $|B_{\mathrm{NV}}|$. The brighter maximum identifies $\phi$. 

\subsection{Experimental Setup and PMA Material}
\label{sec:methods}

All measurements were conducted using a commercial scanning NV microscope (Qnami - ProteusQ). The system utilizes a tuning-fork based atomic force microscopy (AFM) feedback loop to maintain probe-sample contact in a frequency-modulation mode (Set point of \SI{3}\hertz{}). To demonstrate calibration of $d_{\mathrm{NV}}$, we use a Quantilever MX probe (FRO76-8B-F3-1K14) from QNAMI consisting of a tapered diamond nanopillar with a flat (100) surface as apex as shown in Figure~\ref{fig:NVSchematic}(b)]. To explore if we can determine the lateral localization of the NV in the pillar, we additionally investigate Quantilever MX+ probes, which have a parabolic apex, to enhance fluorescence collection. The magnetic sensor is a single NV center hosted within the said MX probe, which is excited with a \SI{515}{\nano\meter} laser. The microwave (MW) excitation required for Optically Detected Magnetic Resonance (ODMR) is delivered via a free-standing near-field antenna (which is a \SI{25}{\micro\meter} diameter Au wire) mounted on an xyz translation stage, positioned close to, but not in mechanical contact with, the diamond probe or the sample. 

The magnetic thin-film stack was deposited on a p-doped silicon substrate with \SI{100}{\nano\meter} thermally grown SiO$_2$. Deposition is performed by magnetron sputtering using a Singulus Rotaris sputtering tool at Johannes Gutenberg University Mainz. The stack was Substrate/Ta(4)/Pt(4)/Co(1)/Ru(2), where the numbers in parentheses denote the nominal layer thicknesses in nanometers. The Ta and Pt layers serve as seed/buffer layers for the magnetic Co layer, with the PMA arising from the the Pt/Co interface, while Ru was used as a metallic capping layer to prevent oxidation \cite{Han2019, Kammerbauer2023}. For calibration, all structures are investigated in their as-fabricated state. 

Following deposition, the micron-sized stripes and discs were patterned from the same film using two approaches: we define etch masks via direct laser writing (positive resist) for the stripes and electron-beam lithography (negative resist) for the discs. In both cases, the pattern was transferred by argon ion-beam etching under identical conditions (etch time: \SI{60}{\second}). Table \ref{tab:Fabrication} summarizes all fabrication parameters.

\begin{figure}[t]
    \centering
    \includegraphics[width=0.8\columnwidth]{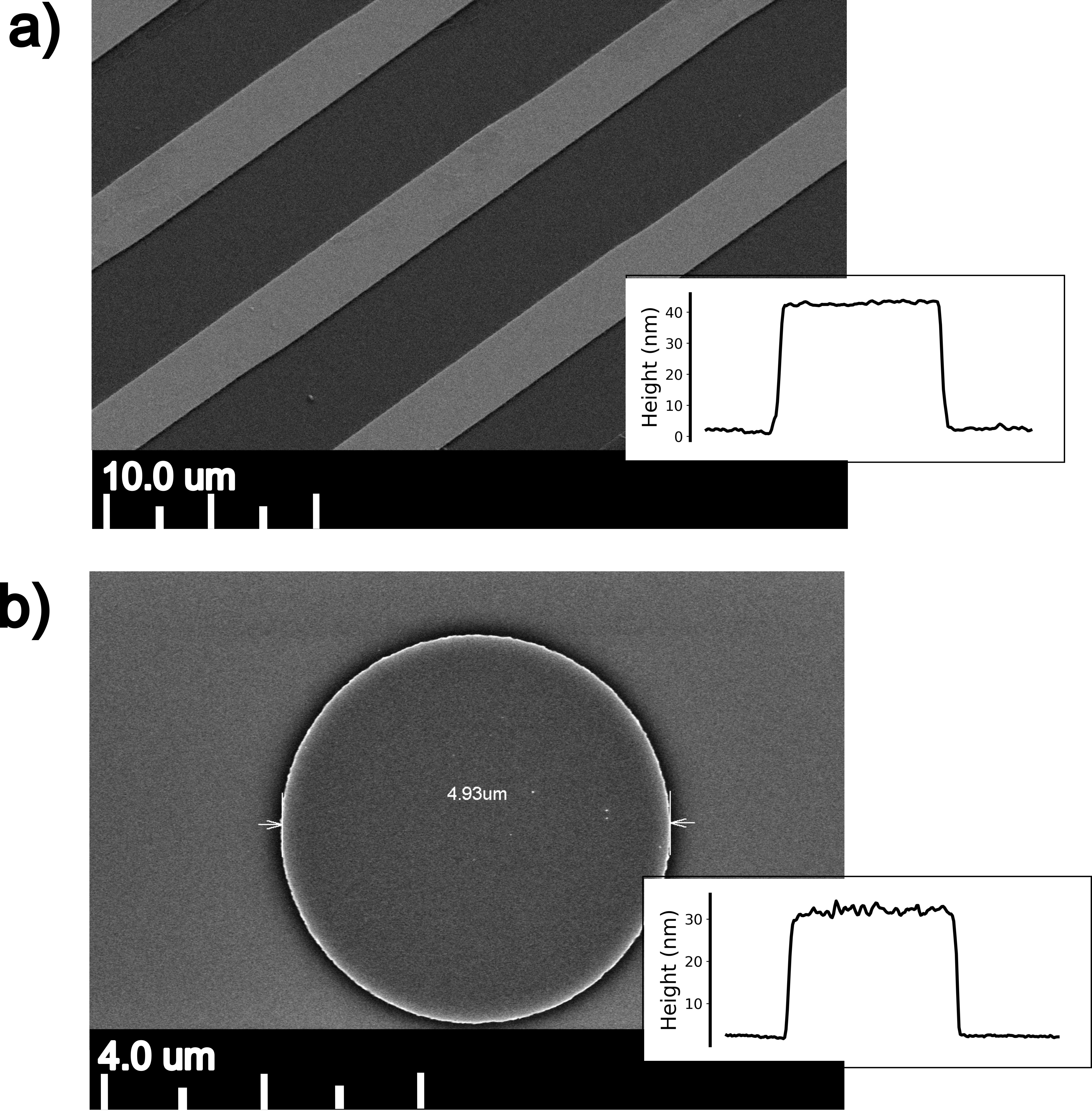}
    \caption{\textbf{(a)} SEM image (main) and AFM topography (inset) of the PMA stripe. 
    \textbf{(b)} SEM image (main) and AFM topography (inset) of the PMA disc}
    \label{fig:SEM}
\end{figure}

The residual resist was removed using Dimethyl Sulfoxide (DMSO). Using AFM, we estimate an edge roughness of $\approx 6 - 12$ \si{\nano\meter}, consistent with the smooth appearance of the stripes in SEM images [Figure~\ref{fig:SEM}]. This roughness is small compared to $d_{\mathrm{NV}}$, and its effect on the extracted $d_{\mathrm{NV}}$ lies below the scan-to-scan scatter. The inset of Figure~\ref{fig:SEM} shows the stripe profile obtained from AFM which shows that we etch about \SI{40}{\nano\meter}, thus also $\approx$ \SI{30}{\nano\meter} of the substrate has been etched. Since the etch rate used was calibrated only for a reference material (chromium $\approx$ \SI{20}{\nano\meter\per\minute}), we chose an etch time of \SI{60}{\second} to guarantee complete removal of the magnetic stack and thus get a magnetically sharp edge. The underlying substrate is non-magnetic, so etching deeper into the bare substrate creates no additional stray field, and the only magnetic source remains the stripe edge itself. The topographic step also does not compromise the extraction of $d_{\mathrm{NV}}$ as explained in the following: First,  we obtain $d_{\mathrm{NV}}$ from the fitting $B_{\mathrm{NV}}$, whose characteristic width is set by the rise and peak of the sharp curve on the film side. Second, the apex of the probe descends onto the substrate only once it has completely passed the edge.  So even when imaging the region directly at the edge, the NV center height is still unchanged. Third, for each line cut, the number of data points collected on the film side is much higher than those collected on the substrate thus further reducing the influence on $d_{\mathrm{NV}}$.

Depending on the measurement, integration times per pixel are chosen to balance signal-to-noise ratio and scan duration. The scans are recorded at a laser power of \SI{0.2}{\milli\watt} and a microwave power of $29$ ~dBm. The integration time per pixel is chosen to be between 30 - 40 \si{\second}. Multiple scans are averaged when required to improve statistical precision. Complete field maps ($100\times10$ pixels for stripes and $40\times40$ pixels for discs) typically required $14$ -- $19$~h to acquire depending on the number of pixels.   For the inverse AFM, we use a commercially available test grating (NT-MDT, TGT1).

\subsection{Calibration Procedures}
\label{sec:protocol}

In this section, we outline the procedure to calibrate $d_{\mathrm{NV}}$ and $\phi$ using the PMA microstructures modeled in Section~\ref{sec:pma}. First, we determine $d_{\mathrm{NV}}$ from stray field imaging on PMA stripes. Subsequently, $d_{\mathrm{NV}}$ enters the evaluation of two-dimensional stray field images of the PMA discs, allowing to unambiguously extract $\phi$.

In detail, we start extracting  $d_{\mathrm{NV}}$ using one-dimensional line-cuts across a PMA stripe edge (measurement geometry see Figure~\ref{fig:models}). We select an isolated, straight edge segment to ensure the one-dimensional approximation $B_y \approx 0$ holds. Scans are performed perpendicular to the edge with a lateral range that captures the full extrema of the stray field components. During the fitting process, we treat the edge position $x_0$, the bias field $B_0$ and  the field amplitude $A = \mu_0 M_s t$ (and thus the effective magnetization $M_\mathrm{s}$) as free parameters alongside $d_{\mathrm{NV}}$. This is a critical step for us, as it accounts for potential offsets between the topographic edge and tiny deviations in magnetization due to film thickness during fabrication, while allowing $d_{\mathrm{NV}}$ to be determined accurately. To ensure good statistical reliability, multiple line-cuts are averaged along the stripe, providing a mean $d_{\mathrm{NV}}$ together with its standard deviation across the line-cuts. This spread arises from local surface roughness and small drifts in the AFM feedback height that might occur over the course of the scan.

\begin{figure}[hb]
    \centering
    \includegraphics[width=1\columnwidth]{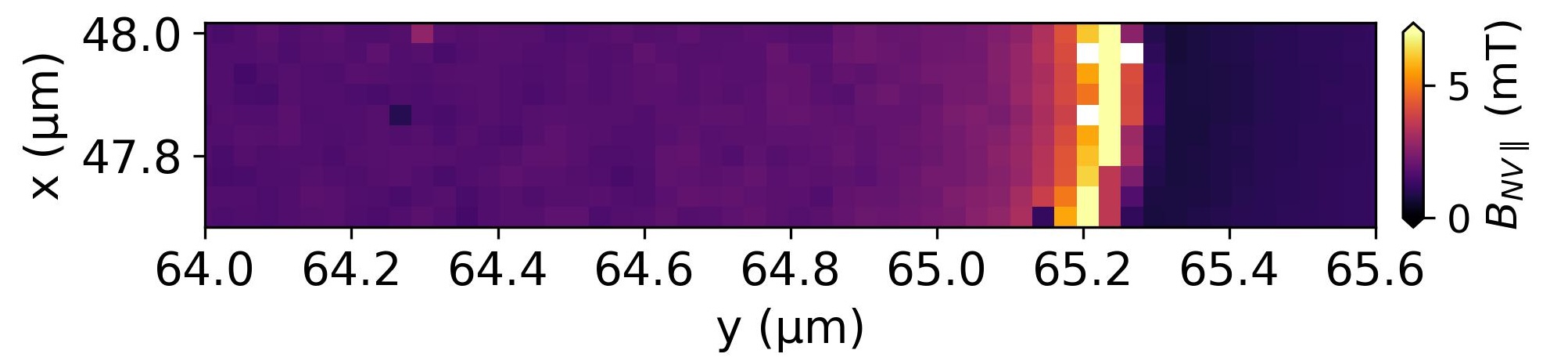}
    \caption{\textbf{Two-dimensional $B_{\mathrm{NV}}$ map over the stripe region.} The map shows the stray magnetic field $B_{\mathrm{NV}}$ measured over the edge of a PMA stripe. The color scale indicates $B_{\mathrm{NV}}$ in \si{\milli\tesla}.}
    \label{fig:field_stripe}
\end{figure}

\begin{figure}[!h]
    \centering
    \includegraphics[width=1\columnwidth]{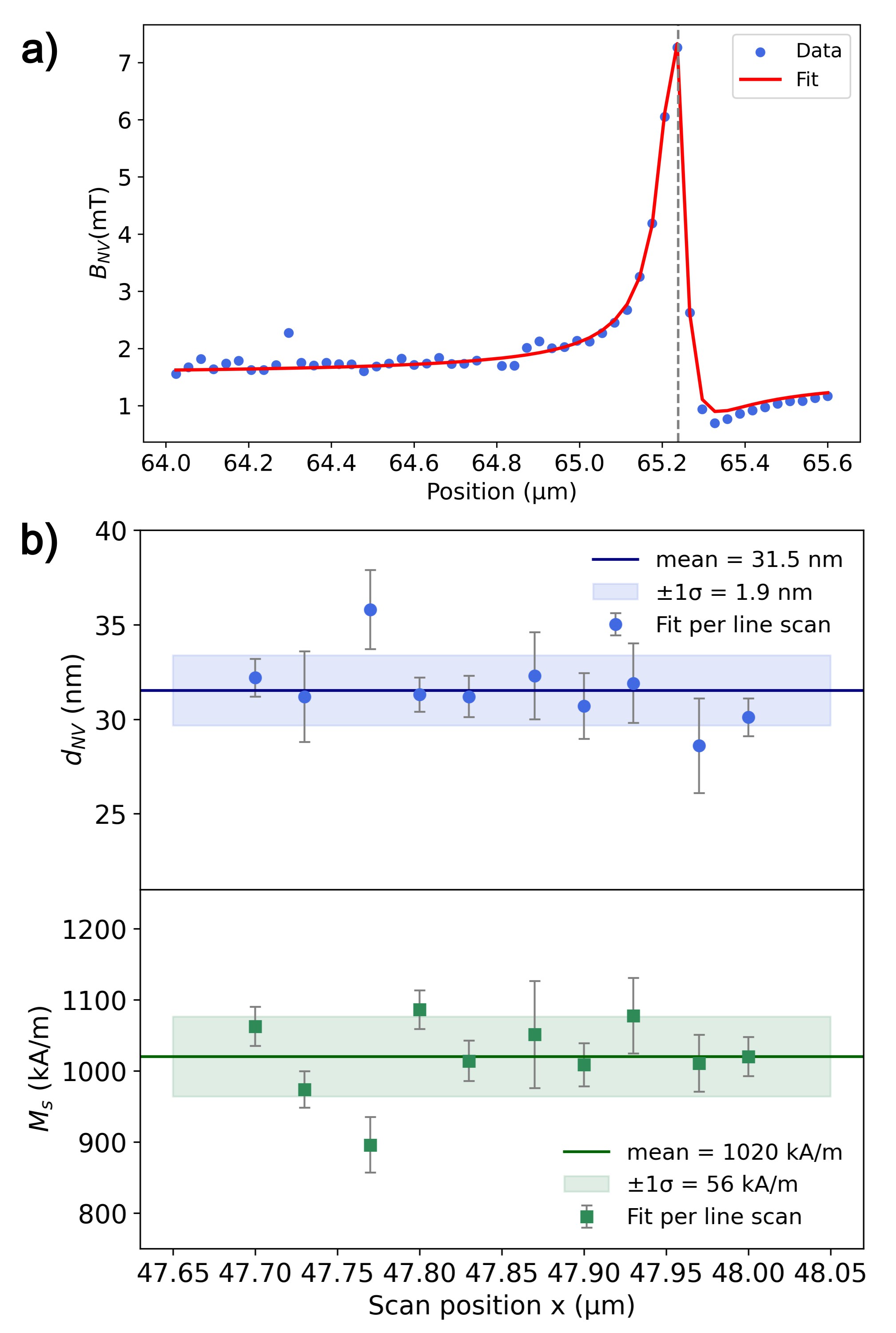}
    \caption{\textbf{NV-to-sample distance calibration from stripe edge line scans.} (a) Representative $B_{\mathrm{NV}}$ line scan (blue points) acquired perpendicular to the stripe edge, together with the fitted stray-field model (red curve). (b) Summary of $d_{\mathrm{NV}}$ (top) and $M_{\mathrm{s}}$ (bottom) values extracted from ten independent line scans distributed along the stripe edge. The solid lines and shaded bands indicate the mean and the $\pm 1\sigma$ standard deviation across the ensemble, respectively. }
    \label{fig:stripe_fit}
\end{figure}

With $d_{\mathrm{NV}}$ fixed, we determine  $\phi$ from a two-dimensional field map of a PMA disc. We fit the two-dimensional map to \eqref{eq:projection} with the stray field given by the current-loop model in equations~\eqref{eq:bz_loop} and \eqref{eq:br_loop}. For every scan, the disc center may not perfectly coincide with the origin of the scan frame (due to sample placement or drift). Consequently, we use the scan center as an initial guess for the center of the disc but treat the center of the disc $(x_0, y_0)$ as a free fitting parameter. $R$ is then the fitted loop radius relative to this center. The free parameters of the fit are therefore the disc center $(x_0, y_0)$, the disc radius $R$, the bias field $B_0$, the current $I$ ( equivalently $\sigma M_s t$ )and the azimuthal angle $\phi$.

The radial field $B_r (r,z)$ of the disc projected onto the NV axis $B_{\mathrm{NV} \phi}(r,\alpha)$ is expected to peak in the direction of $\phi$ and to produce a bright arc on the disc edge. The angular position of this arc, taken alone, gives $\phi$, with a \ang{180} ambiguity that depends on the magnetization direction $\sigma$ of the disc. The same map however also encodes $\sigma$: at the disc center, the stray field is purely axial by symmetry, so it simply adds to or subtracts from $B_0$. Comparing the brightness of the disc's interior to the far background therefore can tell whether the disc field adds to or opposes $B_0$, so fitting the full map allows us to fix $\sigma$ and hence $\phi$ together. The measured angle may then be assigned to one of the four crystallographically allowed $\langle 111 \rangle$ orientations, whose in-plane projections are separated by \ang{90}. 

Together with the polar angle $\theta$, set by the diamond crystal orientation, this procedure fully determines the NV axis $\mathbf{n}_{\mathrm{NV}}$. Once calibrated, the parameters $(d_\mathrm{NV}, \theta, \phi)$ can be used directly in quantitative field reconstruction for subsequent measurements. Recalibration is typically only required upon probe exchange, after mechanical instabilities, or after long term usage of the probes.

\begin{figure*}[!t]
    \centering
    \includegraphics[width=\textwidth]{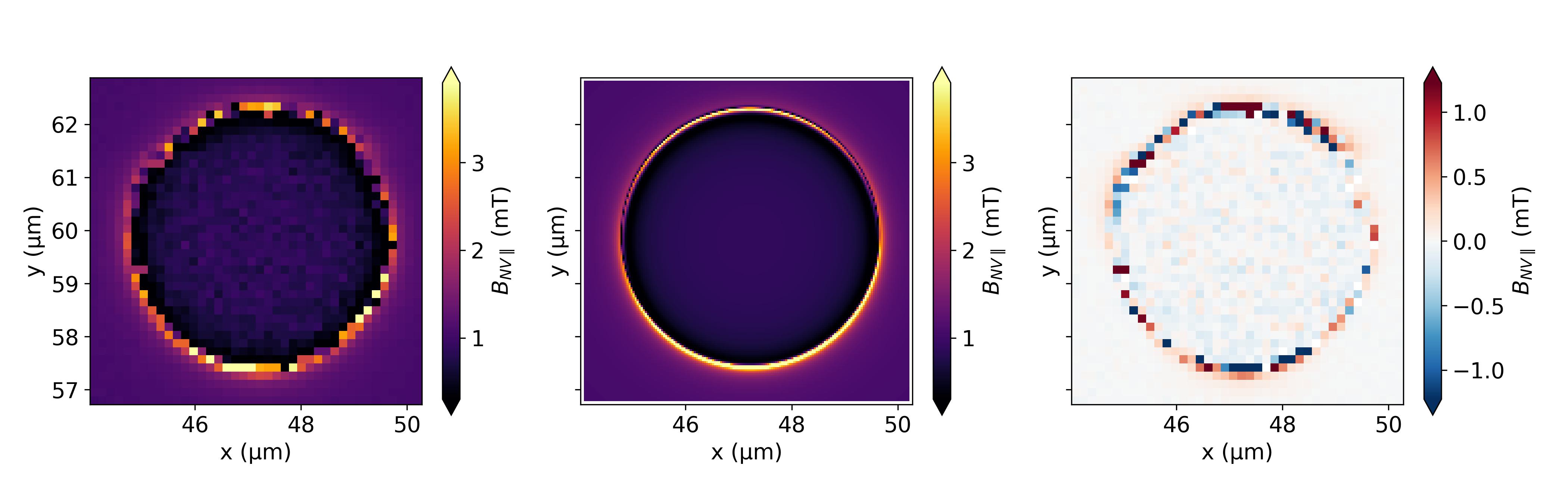}
    \caption{\textbf{NV orientation from a PMA disc} Measured $B_{\mathrm{NV}\parallel}$ map (left), model of Eq. \eqref{eq:projection} evaluated at the best fit parameters (center), and the residual between the two (right) }
    \label{fig:disc_fit}
\end{figure*}

\subsection{NV Localization and Contamination Detection}
To complete the characterization of the diamond nanopillar probe, we perform inverse AFM scanning the diamond nanopillar over an ultra-sharp, commercially available silicon needle. The measured topography consists of a convolution of the diamond nanopillar geometry and the silicon needle geometry. However, the latter is much sharper (radius of curvature $\leq$~\SI{10}{\nano\meter}) compared to the ``blunt'' diamond nanopillar with $\leq$~\SI{200}{\nano\meter} diameter. 

Simulations of the collected NV fluorescence during inverse AFM measurements were performed using Lumerical FDTD \cite{lumerical_fdtd}. The simulation model included both the diamond nanopillar and the silicon needle, together with their respective substrate. The refractive index (n) and extinction coefficient (k) of diamond were taken from Ref.\ \cite{phillip1964diamond}, while the optical constants of silicon were taken from Refs.\ \cite{palik1985handbook, pierce1972amorphoussi}. The NV center was modeled as a broadband electric dipole source with a wavelength range of $600 - 800$ \si{\nano\meter}, positioned within the diamond nanopillar and oriented at ($\theta=$ \ang{53}) with respect to the diamond (100) axis. We here adopt $\theta$ as given by the manufacturer \cite{Qnami_QuantileverMX}, while for the evaluation of the magnetic measurements, we stick to $\theta=$ \ang{54.7} as deduced from the ideal crystal orientation. A mesh override region with a uniform mesh size of \SI{3}{\nano\meter} was applied around both the diamond nanopillar and the silicon needle, while the remainder of the simulation domain used Lumerical's mesh accuracy setting of 3. Further details of the simulation geometry are presented in  Section~\ref{sec:results}.

\section{Results}
\label{sec:results}

\subsection{Distance and orientation calibration}
Figure~\ref{fig:field_stripe} shows a two-dimensional map of $B_{\mathrm{NV}}$, measured across a PMA stripe edge.  $B_{\mathrm{NV}}$ is uniform along the length of the edge, consistent with a homogeneous magnetization and supporting the one-dimensional approximation ($B_y \approx 0$) employed in the analytical model. To evaluate the reproducibility of the calibration, we extract ten linecuts perpendicular to the edge at different positions along the stripe spanning approximately $2.6~\mu$m.

Figure~\ref{fig:stripe_fit}(a) shows a representative line scan and its fitted profile . The stray-field model  in Section~\ref{sec:pma} for a stripe edge reproduces the measured $B_{\mathrm{NV}}$ profile well across the full scan range, with a fitted stand-off distance of $d_{\mathrm{NV}} = 30.7 \pm 1.3$~\si{\nano\meter} for this scan. The quoted uncertainty is the $1\sigma$ confidence interval obtained from the least-squares covariance matrix. Figure~\ref{fig:stripe_fit}(b) summarizes  $d_{\mathrm{NV}}$ as well as the magnetization $M_{\mathrm{s}}$ of the PMA extracted from all ten line scans. We find  a mean  $d_{\mathrm{NV}}$ = \SI{31.5}{\nano\meter} with a standard deviation of \SI{1.95}{\nano\meter}, corresponding to a relative uncertainty of approximately 6\%.  $A = \mu_0 M_s t$ was treated as a free parameter in each fit and the  corresponding $M_\mathrm{s}$ values are consistent across scans, with a mean of $ 1020 \pm 56$\si{\kilo\ampere\per\meter}, in close agreement with the value of $1116$\si{\kilo\ampere\per\meter} from SQUID (superconducting quantum interference device) measurements on the as-grown film. 
We note that treating  $A = \mu_0 M_s t$ as free parameter does not compromise the reliability of determining $d_{\mathrm{NV}}$: $A$ enters the stray field model (equations \eqref{eq:bx_stripe} and \eqref{eq:bz_stripe}) only as an overall amplitude pre-factor. $d_{\mathrm{NV}}$, on the other hand, sets the characteristic width of the profile. Consequently, $d_{\mathrm{NV}}$  and  $A$ can be extracted independently of each other as long as the scan has enough points across the edge so that it captures the full shape of the stray field and not just the peak value. This is further supported by the resulting $M_\mathrm{s}$ values, that are in close agreement with measurement on the as-grown film.

Figure~\ref{fig:disc_fit}(a) shows a two-dimensional map of $B_{\mathrm{NV}}$ for a disc.  Figure~\ref{fig:disc_fit}(b) depicts the field from the model of Equation \eqref{eq:projection} evaluated at best fit parameters. $d_{\mathrm{NV}} =$ \SI{31.5}{\nano\meter} was used as a fixed parameter. The fit for this particular disc returned a disc radius $R =$ \SI{2.43}{\micro\meter}, magnetization $M_\mathrm{s} =$ \SI{997}{\kilo\ampere\per\meter} with $\sigma = -1$, $B_0 =$ \SI{1.03}{\milli\tesla}, and azimuthal angle $\phi=$ \ang{98.6}. In the map, the disc interior appears darker than the far background  - which is consistent with the sign of $\sigma$ the fit returns. A single bright arc appears on the lower edge of the measured disc - this asymmetry being the signature of the NV's azimuth $\phi$. Also, the residual deviation from the fit [Figure~\ref{fig:disc_fit}(c)] is small and mostly unstructured over the bulk of the disc, with some mismatch along the disc rim.

\begin{figure}[t]
    \centering
    \includegraphics[width=0.9\columnwidth]{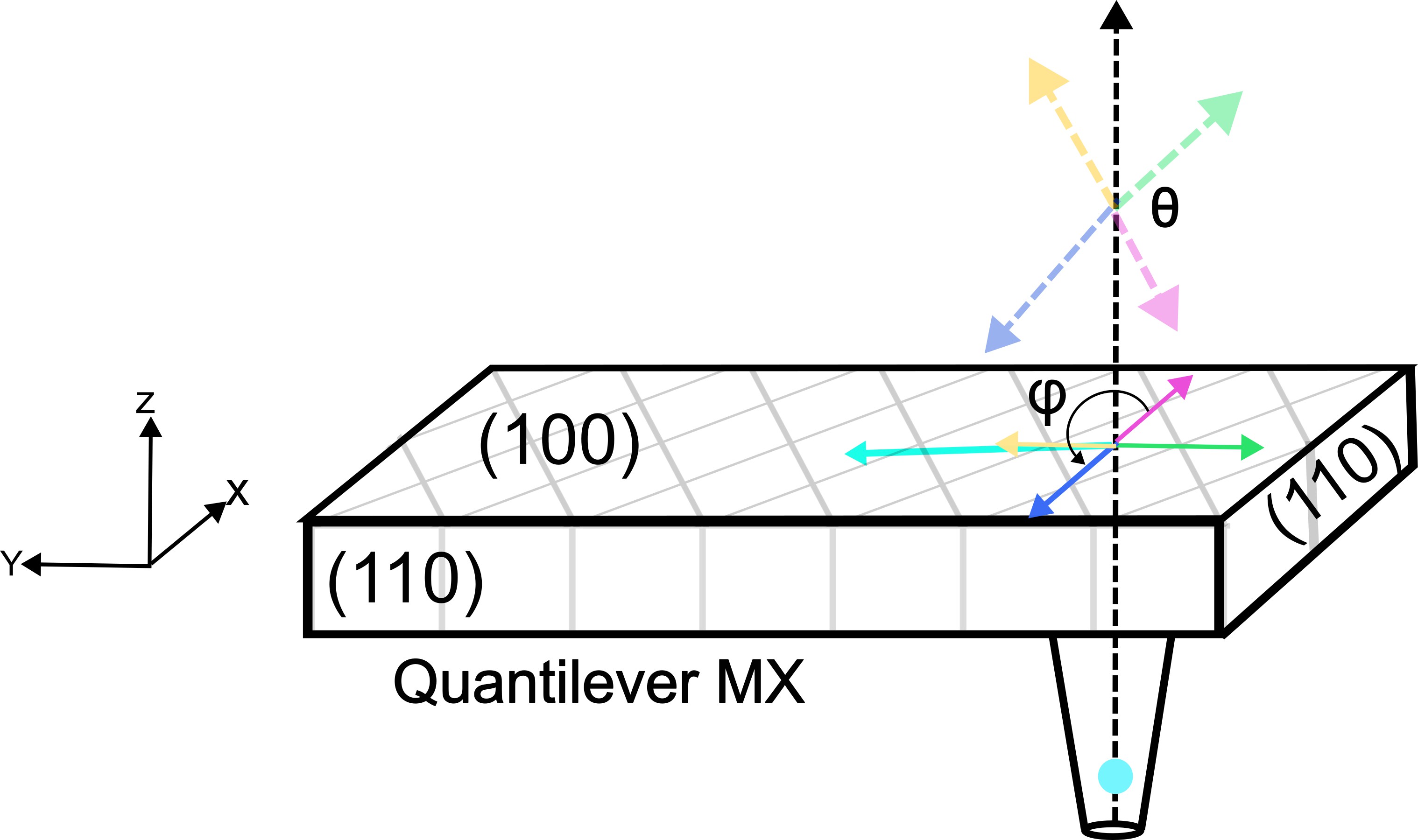}
    \caption{\textbf{Geometry in the lab frame vs. diamond crystal lattice orientations in the probe.} The diagram illustrates the laboratory coordinate axes ($x, y, z$) relative to the physical structure of the commercial diamond probe (Quantilever MX). The top and bottom face of the cantilever-like mounting structure corresponds to the $(100)$ plane, while the elongated side facets are diced along the $(110)$ planes. At the nanopillar apex, the four possible crystallographic $\langle 111 \rangle$ orientations of the NV center are shown as three-dimensional vectors (top arrows) forming a fixed polar angle $\theta = $ \ang{54.7}with respect to the vertical surface normal ($n_{s}$). The corresponding in-plane spatial projections onto the $(100)$ plane form a cross (flat arrows) aligned parallel and perpendicular to the cantilever boundaries.}
    \label{fig:QNAMIphi}
\end{figure}

To assess reproducibility, we repeat the same procedure on two similar discs, giving $\phi=$ \ang{97.3} and $\phi=$ \ang{95.4}. The three values agree to within \ang{3}, confirming that we reliably extract $\phi$ from imaging the discs which together with $\theta=$ \ang{54.7} , helps fully determine the NV axis $\mathbf{n}_{\mathrm{NV}}$ from eq.~\ref{eq:n_nv}.  To evaluate whether the extracted $\phi$ is consistent with the sensor's geometry, we consider the crystallographic orientation of the commercial diamond probe. As schematically illustrated in Figure~\ref{fig:QNAMIphi}: the  $(110)$ side facets of the Quantilever MX inherently lock the four possible NV projections into a ``cross'' (parallel or perpendicular to the cantilever axes). With the long edge of the cantilever-like platform mounted along the $y$ direction in lab frame, the expected $\phi$s are - \ang{0}, \ang{90}, \ang{180}, \ang{270}. Our extracted mean value of $\phi \approx$ \ang{97.1}is in close  proximity to the expected nominal \ang{90} projection (highlighted in Figure~\ref{fig:QNAMIphi}). We analyze the remaining deviation and its relation to  mounting of the sensor assembly, in Section~\ref{sec:discussion}. From the fits, we extract $M_\mathrm{s}= 908 \pm 89$\si{\kilo\ampere\per\meter} which is approximately 11\% lower than the value we found from the measurements of the stripe. This is discussed in detail in Section~\ref{sec:discussion}.

\subsection{NV localization and probe contamination}
\begin{figure}[!h]
    \centering
    \includegraphics[width=\columnwidth]{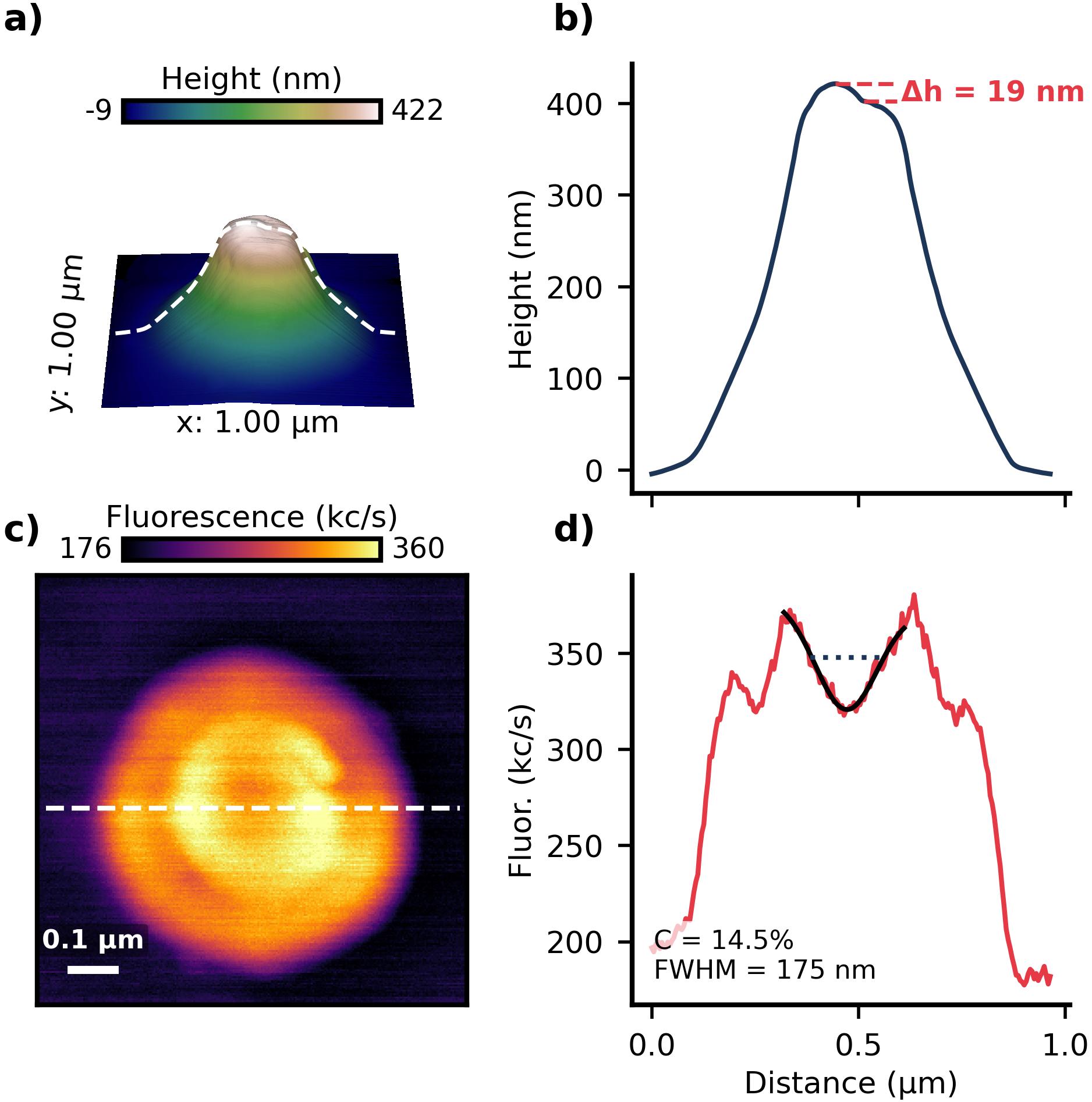}
    \caption{Inverse AFM of the conical MX probe FRO76-8B-F3-1K14 used for the depth and orientation measurements. 
    (a) Three-dimensional topography displayed in its physical aspect ratio, featuring a profile path (white dashed line) through the center which is plotted in panel (b). The step height $\Delta h$ of a potential contamination at the bottom surface of the diamond pillar is highlighted in red. 
    (c) Two-dimensional fluorescence map recorded simultaneously, with the corresponding line cut displayed in panel (d). A drop in fluorescence is visible in the center, fitted using an inverted Gaussian function with a linear background (solid black line). The contrast $C = 14.5\%$ and the FWHM linewidth of \SI{175}{\nano\meter} (dotted blue-gray line) are stated in the bottom left inset.}
    \label{fig:Inverse_AFM_1K14}
\end{figure}

\begin{figure}[!h]
    \centering
    \includegraphics[width=\columnwidth]{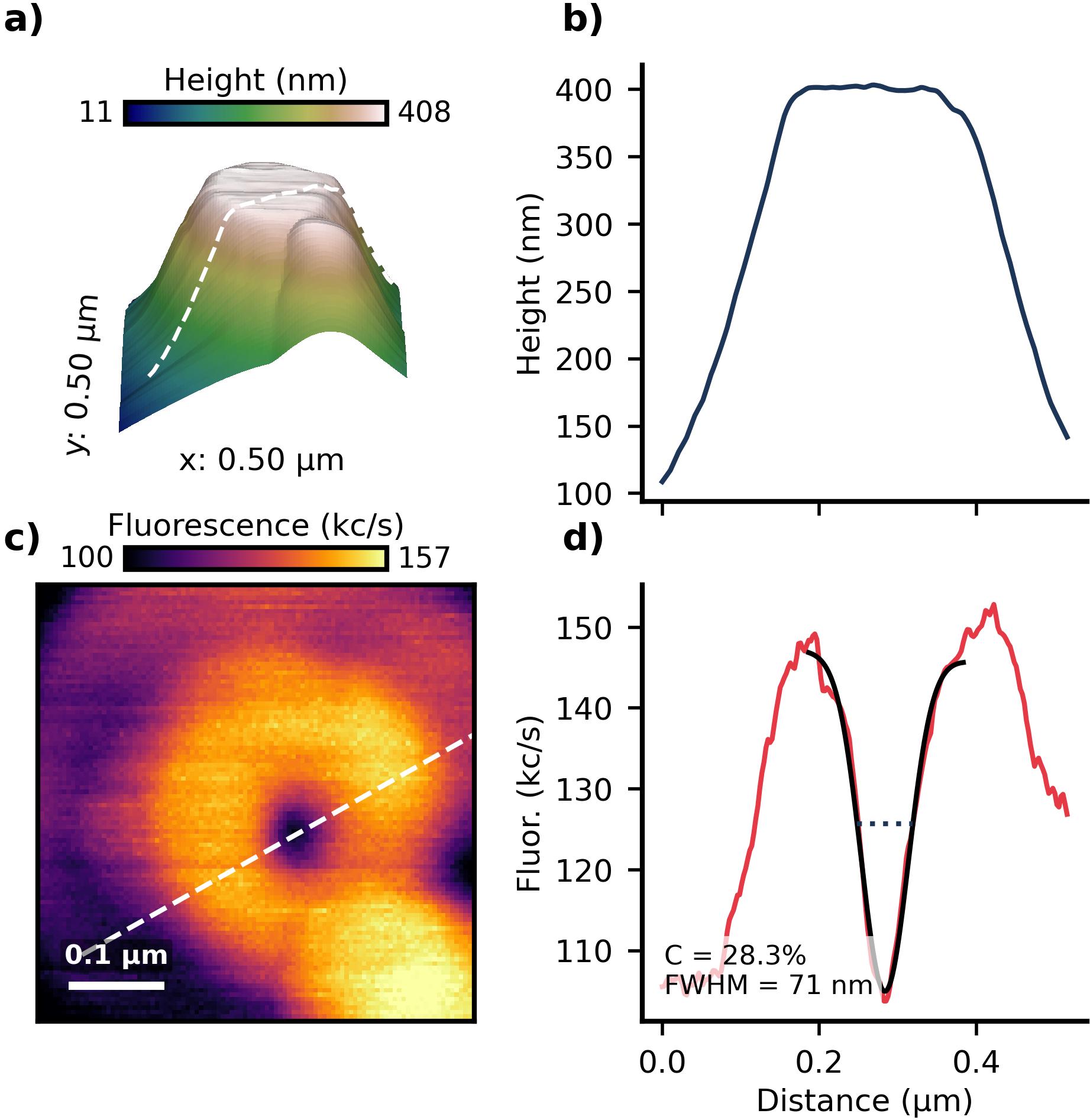}
    \caption{Inverse AFM of a conical MX probe featuring a particle firmly attached to the side of the tip. 
    (a) Three-dimensional topography displayed in its physical aspect ratio, showing the conical shape of the diamond pillar with a lateral particle attachment. An oblique profile path, chosen to avoid distortion from the attached particle, is indicated by the white dashed line and plotted in panel (b). 
    (c) Two-dimensional fluorescence map recorded simultaneously, with the corresponding line cut displayed in panel (d). A sharp fluorescence dip is visible in the center, fitted using an inverted Gaussian function with a linear background (solid black line). The contrast $C = 28.3\%$ and the FWHM linewidth of \SI{71}{\nano\meter} (dotted blue-gray line) are stated in the bottom left inset.}
    \label{fig:Inverse_AFM_1F14}
\end{figure}

\begin{figure}[!h]
    \centering
    \includegraphics[width=\columnwidth]{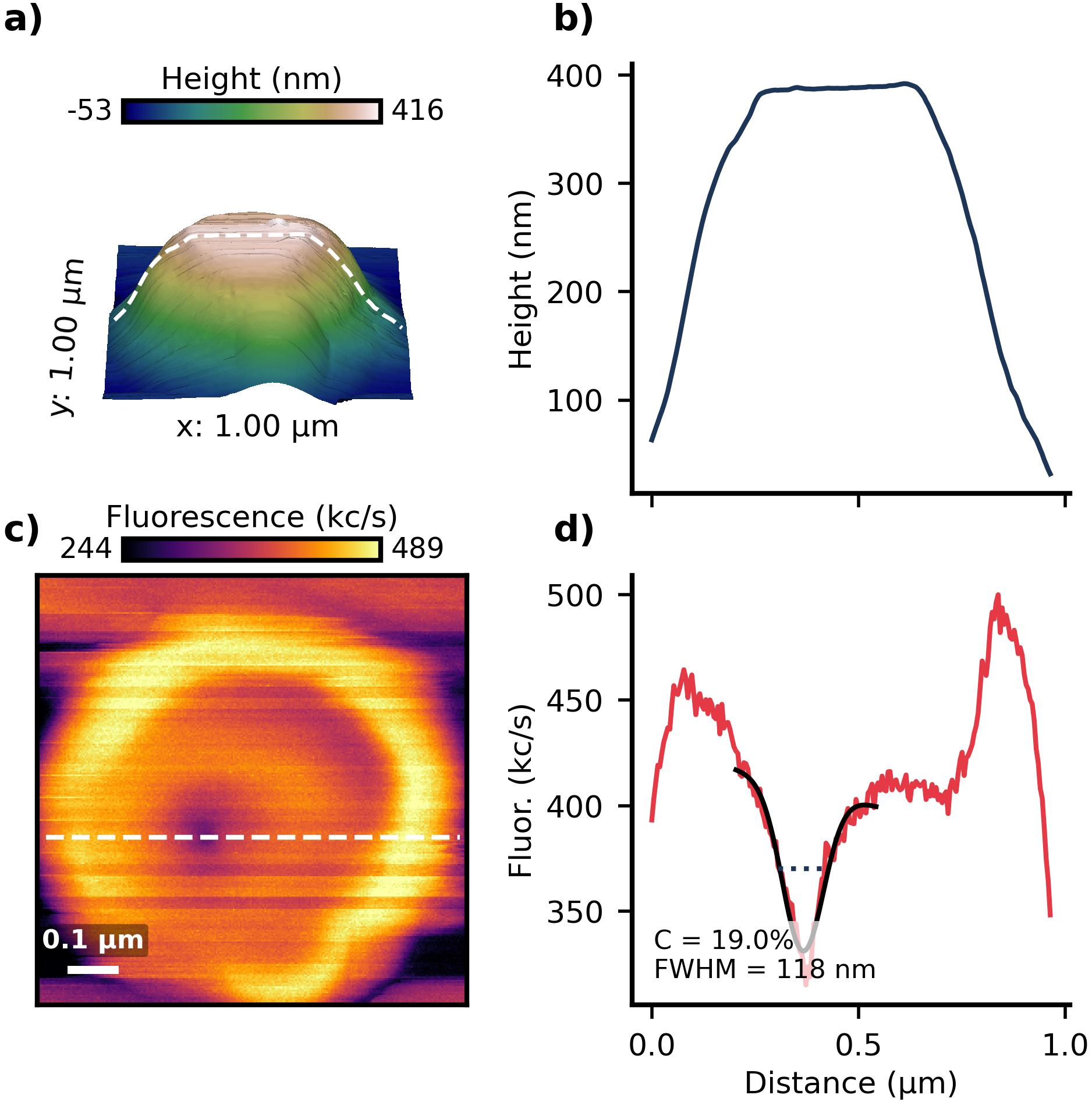}
    \caption{Inverse AFM of a parabolic MX+ probe. 
    (a) Three-dimensional topography displayed in its physical aspect ratio. A profile path is indicated by the white dashed line and plotted in panel (b). 
    (c) Two-dimensional fluorescence map recorded simultaneously, with the corresponding line cut displayed in panel (d). A fluorescence dip is visible off-center, fitted using an inverted Gaussian function with a linear background (solid black line). The contrast $C = 19.0\%$ and the FWHM linewidth of \SI{118}{\nano\meter} (dotted blue-gray line) are stated in the bottom left inset.}
    \label{fig:Inverse_AFM_D11}
\end{figure}

In this section we turn to evaluating the lateral position of the NV center within the diamond pillar. We image the probe using inverse AFM scans across a \SI{1}{\micro\meter} $\times$ \SI{1}{\micro\meter} region (\SI{0.5}{\micro\meter} $\times$ \SI{0.5}{\micro\meter} for Figure~\ref{fig:Inverse_AFM_1F14}) centered around the silicon needle, with a step size of \SI{4}{\nano\meter} (\SI{5}{\nano\meter} for Fig.~\ref{fig:Inverse_AFM_1F14}). In Figure~\ref{fig:Inverse_AFM_1K14} and Figure~\ref{fig:Inverse_AFM_D11}, a triangular tilt correction defines the flat surface of the silicon needle grating as the zero plane, and the topography data in Figure~\ref{fig:Inverse_AFM_1K14} is corrected for line artifacts by aligning the start of every scan line.

For this experiment, we first use the same diamond nanopillar probe employed in calibrating $d_{\mathrm{NV}}$ via magnetic field scans on the PMA structures. The topography for this probe is shown in Figure~\ref{fig:Inverse_AFM_1K14}(a) with its corresponding line cut displayed in panel (b), while the simultaneously recorded two-dimensional fluorescence map and its cross-section are presented in panels (c) and (d), respectively. 

We extract two key findings from this initial measurement. First, the topography allows us to evaluate the contamination state of the probe, as particles attached to the tip become visible. A distinct feature at the very top of the topography cross-section, adding approximately \SI{19}{\nano\meter} in height, points to contamination at the flat bottom surface. Second, the detected NV center fluorescence rate drops when the silicon needle is positioned close to the center of the nanopillar. 
While the overall fluorescence landscape can be explained by topographic features, this central dip is attributed to the close proximity of the needle to the NV center, as discussed in Ref.~\cite{Maletinsky2012} for a metalized needle. In Section \ref{sec:Numerical}, we investigate this effect using numerical simulations, verifying that this conclusion is justified and showing that the dip is also expected without a metallic coating on the silicon needle.
The full-width-half-maximum (FWHM) linewidth of this dip is \SI{175}{\nano\meter} with a contrast of approximately \SI{14.5}{\percent}, extracted from a Gaussian-dip fit with a linear background of the central region shown in Figure~\ref{fig:Inverse_AFM_1K14}(d).

To highlight how features can vary between individual probes, Figure~\ref{fig:Inverse_AFM_1F14} shows the inverse AFM scan of another conical MX probe. Contamination is evident on the side of this pillar, which in this case does not affect the magnetic stand-off distance $d_{\mathrm{NV}}$. To avoid artifacts from this lateral particle, the profile path [white dashed line in panels (a) and (c)] is intentionally positioned obliquely, as the contamination clearly impacts the surrounding fluorescence landscape shown in panel (c). 

A highly pronounced fluorescence drop is visible along this oblique line cut in panel (d). The FWHM linewidth of the dip is \SI{71}{\nano\meter} and the contrast is approximately \SI{28}{\percent}, extracted from the corresponding inverted Gaussian fit with a linear background. Notably, this contrast is roughly twice as strong as that observed for the previous probe in Fig.~\ref{fig:Inverse_AFM_1K14}(d), while the FWHM linewidth is significantly narrower.

An additional inverse AFM scan of a different type of diamond nanopillar (MX+ type, parabolic shape) is presented in Figure~\ref{fig:Inverse_AFM_D11}. The three-dimensional topography in panel (a) as well as the line cut in panel (b) underline the structural differences between both probe designs [compare Fig.~\ref{fig:Inverse_AFM_1K14}(a), (b)], which must be taken into account when interpreting topographic data. The two-dimensional fluorescence map in panel (c) shows a local minimum which, in contrast to Figure \ref{fig:Inverse_AFM_1K14}, is positioned off-center. The Gaussian fit in panel (d) confirms that the fluorescence minimum is indeed shifted by approximately \SI{80}{\nano\meter} from the center of the nanopillar. This directly demonstrates that the lateral NV location can vary significantly from probe to probe. Furthermore, this specific measurement appears noisy, exhibiting sharp jumps in the fluorescence intensity as well as horizontal disruptions in the topography (which is shown without line correction, as the origin is not thermal drift in this case). We attribute this effect to a loosely attached particle that escapes direct imaging by the inverse AFM, but is instead pushed around by the silicon needle during scanning.

\subsection{Numerical simulations}
\label{sec:Numerical}
\begin{figure}[!h]
    \centering
    \includegraphics[width=\columnwidth]{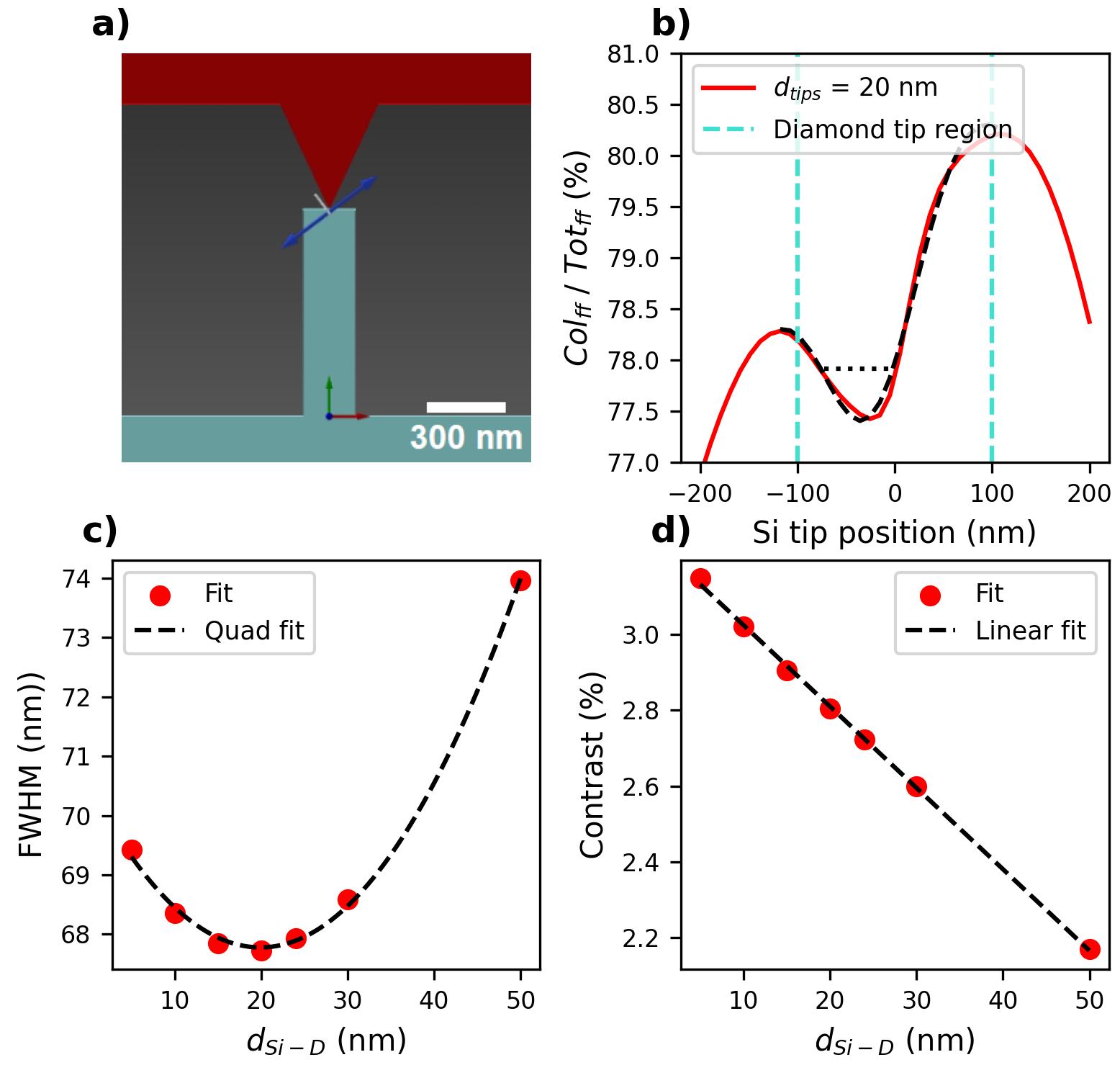}
    \caption{\textbf{Simulation of the inverse AFM measurements.} (a) Two-dimensional model in the $xy$ plane showing the silicon needle (red), diamond nanopillar (turquoise), and dipole approximating the NV center (blue arrows). (b) Simulated results  for ($d_\mathrm{{Si-D}}$= \SI{20 }{\nano\meter}). The red curve shows the fraction of collected farfield emission  ($Coll_\mathrm{ff} / Tot_\mathrm{ff}$) obtained by diving the collected farfield emission $Coll_\mathrm{ff}$ by the total farfield emission $Tot_\mathrm{{ff}}$ as a function of the silicon needle's lateral position. Turquoise dashed lines indicate the nanopillar boundaries, the black dashed curve is a Gaussian fit including a third order polynomial, and the dotted line marks the FWHM (\SI{67.7}{\nano\meter}). (c–d) FWHM and contrast obtained from the Gaussian fits (in red) described in the text. }
    \label{fig:lumerical}
\end{figure}

To better understand the origin of the dip observed in the fluorescence scans of the silicon needle using the NV-diamond probe, we perform numerical simulations using the geometry  in Figure~\ref{fig:lumerical}(a).
The silicon needle grating, shown in red, consists of a conical tip with an apex approximated by a flat surface of \SI{10}{\nano\meter} diameter. According to manufacturer specifications, we use an opening angle of (\ang{50}) and a height of \SI{400}{\nano\meter} for the cone. The diamond device (turquoise) includes the diamond nanopillar and mounting structure (``cantilever''). As the manufacturer does not specify the exact geometry of the diamond nanopillar, we approximate it as a cylinder with a diameter of \SI{200}{\nano\meter} and a height of \SI{800}{\nano\meter}.  A frequency-domain field monitor (DFT monitor) below the diamond cantilever records the electric field at \SI{637}{\nano\meter}  to calculate both the total far-field radiation and the fraction collected by our objective with a numerical aperture of NA = 0.7.
We perform two-dimensional simulations in order to allow for time-efficient simulations and to reproduce line profiles comparable to those measured experimentally in Figure~\ref{fig:Inverse_AFM_1K14} (d). To generate these profiles, the silicon needle moves laterally from \SI{-200}{\nano\meter} to \SI{200}{\nano\meter} while maintaining a constant distance $d_\mathrm{{Si-D}}$ to the diamond nanopillar's surface. At each lateral position, we compute the far-field intensity at \SI{637}{\nano\meter}, i.e.\ the position of the NV zero-phonon-line. Figure~\ref{fig:lumerical}(b) shows a representative simulation for $d_\mathrm{{Si-D}} = $ \SI{20} {\nano\meter} . The simulated profile exhibits a pronounced dip. To analyze this dip, we use the approach that has been successfully applied to model our experimental data: a inverted Gaussian function, here with the``background'' approximated using a third order polynomial function. The fitted minimum is located \SI{31.5}{\nano\meter} from the dipole's position (defined as $x = 0$). We note that the measured profiles are more symmetric than the simulated ones. We assume that in the simplified model of a single, tilted dipole, an asymmetric curve is created. In contrast, NV centers are fully characterized by two equally strong dipoles in the plane perpendicular to $\mathbf{n}_{\mathrm{NV}}$ which might lead to a  more symmetric reaction.~\cite{Neu2014, Alegre2007} 
The simulations were repeated for $d_\mathrm{{Si-D}}$  ranging from 5 to 50 \si {\nano\meter}. We note that $d_\mathrm{{Si-D}}=$ \SI{20} {\nano\meter} would correspond to  $d_{\mathrm{NV}} =$ \SI{30}{\nano\meter} which is the value we extracted in the magnetic measurements (assuming that the NV is placed $d_{\mathrm{NV_S}}= $ \SI{10} {\nano\meter} below the diamond surface). Figures \ref{fig:lumerical}(c) and (d) summarize the analysis of the dips. In Figure~\ref{fig:lumerical}(c), the FWHM values of the calculated Gaussian dips as a function of $d_\mathrm{{Si-D}}$ are fitted with a quadratic function, yielding a minimum at $d_\mathrm{{Si-D}} = \SI{19.9}{\nano\meter}$. Figure~\ref{fig:lumerical}(d) shows that the contrast of the Gaussian dips decreases linearly with increasing $d_\mathrm{{Si-D}}$. 
These simulations provide first insights into the origin of the fluorescence dip observed experimentally during scans of the silicon needle, such as those shown in Figure~\ref{fig:Inverse_AFM_1K14}(d). As the silicon needle approaches the NV center, the emitted fluorescence progressively decreases, resulting in a reduction of the collected far-field intensity. The fluorescence reduction is stronger at reduced $d_\mathrm{{Si-D}}$. We consequently associate it with fluorescence quenching attributed to placing the absorbing silicon needle's tip (extinction coefficient $k = 0.019$) in the optical near-field of the NV's dipole. Additionally, the presence of the high refractive index needle in the near-field of the dipole potentially alters its emission profile and consequently the coupling to the photonic modes of the nanopillar. To determine which of the two mechanisms contributes most strongly to the fluorescence reduction at $d_\mathrm{{Si-D}}=$ \SI{20} {\nano\meter}, we performed further simulations using different values of the complex refractive index (n,k) of the silicon needle: (3.88, 0), (3.88, 0.019) and (3.88, 1.9). The difference in the Gaussian-dip contrast between k = 0 and the actual extinction coefficient of silicon, k = 0.019, is negligible. Even when increasing the extinction coefficient by two orders of magnitude to k = 1.9, the contrast increases by only 8\% compared with the lossless case. These results indicate that absorption by the silicon needle contributes only weakly to the observed fluorescence reduction, which is instead predominantly attributed to the effect of its high refractive index on the emission profile of the NV center and its coupling to the photonic modes of the nanopillar.
Our simplified two-dimensional model successfully reproduces the qualitative behavior and allows us to justify that the position of the dip in the collected fluorescence indicates the NV's lateral position. Our simulations, however, do not quantitatively match the measured dip characteristics. In particular, simulated contrast is lower than the experimental value even at the smallest $d_\mathrm{{Si-D}}$, and the simulated FWHM is consistently smaller than that measured experimentally. These discrepancies are likely due to the simplified geometry adopted in the simulations, as well as the simplified model for the NV center. The model could be improved by incorporating the curvature of needle and nanopillar apex, a more realistic diamond nanopillar geometry, and a more accurate description of the NV center position and dipole orientation.

\section{Discussion and Conclusion}
\label{sec:discussion}
We here present methods to calibrate the NV to sample distance $d_{\mathrm{NV}}$ as well as the NV's high symmetry axis orientation especially the azimuthal angle $\phi$. We use standard cleanroom procedures to pattern stripes and discs of a PMA material. Especially, calibration of $d_{\mathrm{NV}}$ can be repeated between different measurements. According to manufacturer specification, NV centers in the probes have a nominal depth $d_{\mathrm{NV_S}}=$ \SI{10}{\nano\meter}. We find a mean value of $d_{\mathrm{NV}}$ = \SI{31.5}{\nano\meter} for the investigated diamond probe, thus clearly highlighting the importance of the additional stand-off distance $d_{\mathrm{AFM}}$ and the need for reliable calibration.  The measured value of $\phi$ is comparable to $\phi= $ \ang{90} which is  one of the four possible orientations we expect from the NV sensor geometry [Figure~\ref{fig:QNAMIphi}] if the [110] edges of the diamond sensor device are perfectly aligned with the y-axis of our sample coordinate system. However, as the diamond probes are attached to a tuning fork via a gluing process to allow for AFM-like feedback, this induces a misalignment that can vary between probes. For the probe used here, a clear misalignment is visible in the optical microscope we use to align the probes with respect to the sample. It thus becomes clear that $\phi$  cannot be inferred from geometry considerations only and needs to be reliably calibrated at least once per probe. 

Beyond sensor geometry, the fits yield a saturation magnetization of $M_\mathrm{s} = 908 \pm 89$\si{\kilo\ampere\per\meter} for the discs which is about $11\%$ lower than the stripes' $1020 \pm 56$\si{\kilo\ampere\per\meter}. This difference may be attributed to several reasons: the small \SI{5}{\micro\meter} discs have a relatively higher perimeter-to-area ratio than the stripes that span through the full length of the wafer, making the discs more sensitive to fabrication damage. For a very thin magnetic layer, this damage may lead to a localized loss of magnetic volume. And because the discs are bound in all directions, the etching process may reduce their magnetic signal more significantly than for continuous stripes which have more ``undisturbed'' film, resulting in a lower $M_\mathrm{s}$. Also, etch masks for the discs are patterned by e-beam lithography while the mask for the stripes are direct laser written. The thicker mask on the stripe might protect the material more efficiently and laser writing might induce less damage than the electron beam lithography.  

We note that the calibration routine presented here does not rely on availability of a vector magnet. Moreover, we demonstrated the routines in ambient conditions but they are compatible with cryogenic operation (given that the sample chamber has enough space to add the calibration sample). 

In this context, our measurements using inverse AFM highlight the importance of contamination control of diamond-based scanning probes: while the diamond and the NV center therein can be used for very long time (the manufacturer quotes typical usability of 6 months), probe contamination might render probes unusable much faster.  It also highlights that especially for a calibration sample, clean surfaces are of highest importance. 
Based on the simulations discussed in Section \ref{sec:Numerical}, we propose that the fluorescence dip observed in the inverse AFM scans can be utilized to determine the lateral position of the NV center within the diamond nanopillar. Furthermore, the dependence of the dip linewidth and contrast on the NV implantation depth (Fig.~\ref{fig:lumerical}) enables a qualitative comparison between individual probes. Therefore, we conclude that the pronounced dip contrast and narrow linewidth measured for the NV center in Fig.~\ref{fig:Inverse_AFM_1F14} indicate a shallowly implanted NV center.

\section*{Funding and acknowledgments}
This work was funded by the Deutsche Forschungsgemeinschaft
(DFG, German Research Foundation) under project number TRR 173–268565370, Spin+X (Project A12).

The PMA sample fabrication and associated material characterization were supported in part by the European Research Council (ERC) under the European Union's Horizon 2020 research and innovation programme (Grant No. 856538, project “3D MAGiC”) and by the TOPOCOM project, funded by the European Union's Horizon Europe Programme Horizon.1.2 under the Marie Skłodowska-Curie Actions (MSCA), Grant Agreement No. 101119608.

We acknowledge the use of the Scanning NV Magnetometer at RPTU, funded by the Deutsche Forschungsgemeinschaft (DFG, German Research Foundation) - 491229782 within the major instrumentation initiative “Spin-based quantum light microscopy (SQLM)”.

We also thank Shaktiranjan Mohanty, postdoctoral researcher at JGU Mainz, for performing the measurements of $M_\mathrm{s}$ for the as-grown PMA films.  

\section*{Data availability}
The data supporting this work is available publicly on Zenodo at DOI: 10.5281/zenodo.21396798

\bibliographystyle{unsrt}
\bibliography{references_v2}

\end{document}